# HETERONUCLEAR DIATOMICS IN DIFFUSE AND TRANSLUCENT CLOUDS

## *T. Weselak*

Institute of Physics, Kazimierz Wielki University, Weyssenhoffa 11, 85-072 Bydgoszcz, Poland;
e-mail: towes@gazeta.pl

**Abstract**

Diffuse and translucent molecular clouds fill a vast majority of the interstellar space in the galactic disk being thus the most typical objects of the Interstellar Medium (ISM).
Recent advances in observational techniques of modern optical and ultraviolet spectroscopy led to detection of many features of atomic and molecular origin in spectra of such clouds. Molecular spectra of heteronuclear diatomic molecules, ie. OH, $OH^+$, CH $CH^+$, CN, NH, CO play an important role in understanding chemistry and physical conditions in environments they do populate. A historical review of astronomical observations of interstellar molecules is presented. Recent results based on visual and ultraviolet observations of molecular features in spectra of reddened, early type OB-stars are presented and discussed.
Appearance of vibrational-rotational spectra with observed transitions based on high-quality spectra, are also presented. Relations between column densities of heteronuclear diatomics (based on the recommended oscillator strengths) and intensities of diffuse interstellar bands (DIBs) are also presented and discussed.

## Introduction

Detection of first molecules in the interstellar medium (ISM) showed that space between the stars despite of being of density lower than that achieved in the best laboratory vacuum is not empty. Molecular bands of simple polar radicals; CH, CN, $CH^+$ have been discovered and identified long ago [1, 2]. For a long period they were believed to be the only possible interstellar molecules. In 1970-ties also homonuclear molecules ($H_2$, $C_2$) have been found ([3, 4]). At the same time rotational emission features revealed the presence of many complex molecules with a dipole momentum in the ISM; the full list of these polar species (mostly carbon-bearing) contains currently nearly 160 entries and demonstrates clearly rich carbon chemistry in interstellar clouds (see [5]). Since the first detections of molecular species astronomers discovered, mostly by means of microwave observations, many diatomics together with their anions and cations, as well as a number of polyatomics in the ISM and still unidentified large species that are the carriers of diffuse interstellar bands (DIBs).

Molecules can be identified through electronic, vibrational and rotational spectra observed through in different spectral regimes. Generally, electronic transitions of simple diatomic molecules arise in the ultraviolet (UV) or visible part of the spectrum; vibrational bands lie at infrared (IR) wavelengths while rotational lines are observed at radio wavelengths [6]. The electronic transitions can be observed using terrestrial spectrographs mounted on big enough telescopes to obatin high enough signal-to-noise and resolving power of the acquired spectra.

In addition to gas-phase molecules, solid dust grains (the total mass of which is typically equals to 1% of the cloud mass [7] play an important role in controlling the physics and chemistry of the cloud in which molecules exist. Interstellar clouds are generally classified on the basis of models and total visual extinction $(A_V)$[1]. In the case of a diffuse cloud such as that observed towards ζ Oph the $A_V$ is equal to 1 mag, whereas in the case of translucent clouds the total visual extinction $A_V$ is expected to be in the range 2-5 mag ([5]). In the case of translucent clouds spectroscopic observations can be performed as still they are optically thin enough to allow radiation from a bright background star to penetrate the cloud. This penetration may trigger photoprocessing which plays an important role in the cloud chemistry. On the other hand, a translucent cloud is not thick enough to make possible radio observations.

In the case of dense clouds, as in Taurus Molecular Cloud (TMC) with $A_V > 10$, very complex molecules can be observed using radio methods, such as $HC_{11}N$ ([8]). The largest currently known interstellar molecules are the fullerenes $C_{60}$ and $C_{70}$ ([9]).

Quite complex chemical species, including large unsaturated carbon chain radicals, fullerenes and prebiotic molecules, have been (and constantly are) synthesized in the ISM and are preserved in these seemingly hostile conditions. Currently available data on abundances of interstellar species are very scarce; it is thus necessary to observe more objects with a much better S/N. The necessary observations are very time-consuming and require well-designed procedures of data acquiring and reducing.

**Observed transitions of heteronuclear diatomics**

Interstellar CH (methylidyne) was identified by McKellar [1] in spectra of several stars of O and B spectral type. The strongest interstellar line of the CH A-X (0, 0) transition centered near 4300 Angstroms (Å) (Fig.1a) is easily observed in the violet region suitable for photographic observations (see Table 1). The CH molecule abundance is known to correlate very well with that of molecular hydrogen and may be used as an $H_2$ tracer ([10]) because the $H_2$ band is situated in the far-UV and thus accessible only to a few dedicated satellites.

The $CH^+$ radical was first identified by Douglas and Herzberg [2] in the ISM. Its strong absorption feature of the A-X (0, 0) transition at 4232 Å is easily seen in Fig. 1b, while interstellar line of the A-X (1, 0) transition at 3957 Å is presented in Fig 2b.
Formation and existence of $CH^+$ in the ISM remains an unsolved problem [5] since the reaction $C^+ + H_2 \rightarrow CH^+ + H$ is endothermic by 4650 K and does not proceed at ISM temperatures which are typically of 100 K. As a result, models such as that of van Dishoeck and Black [5], underpredict the $CH^+$ abundance by about two orders of magnitude.

Absorption features of CN molecule were identified by McKellar [1]. Presented in Fig. 1c spectra show R(0), R(1) and P(1) lines; the first and strongest R(0) line comes from the lowest rotational level, N=0 and R(1) and P(1) lines with N=1 are weaker. The presence of absorption features from more than one rotational level makes it possible to obtain the number of molecules in the ground and first excited level and then to determine the excitation temperature of the CN molecule [11]. It is widely assumed that the CN molecule is in radiative equilibrium with the cosmic microwave background (CMB) radiation in the ISM. The observed excitation temperatures in CN, in fact, do exhibit an excess over the temperature of the CMB which is equal to 2.725 ± 0.002 K [12, 13 and references there in].

In near-infrared spectral region strongly contaminated with telluric lines the CN A-X (2, 0) transition can be observed (see Table 1). In Fig. 3 the spectrum of HD 73882 obtained with MIKE/MAGELLAN instrument is presented.

The OH radical is the first interstellar molecule detected through radio observations [14]; its electronic transitions can be observed at 1222 Å and 3078 Å. The far-UV band was detected using

---

[1] Total visual extinction $A_V=R_V*E(B-V)$, where $R_V$ is typically equal to 3.1 and E(B-V) is the colour excess.

Copernicus Satellite by Snow [15] and the near-UV band (see Table 1 and Fig. 4) by Crutcher and Watson [16].

Until this compilation the OH molecule was detected toward 16 translucent sightlines [17]. Microwave absorption due to the OH in the ISM was observed by Crutcher [18] and Liszt and Lucas [19]. Based on these observations column densities of the OH radical are generally consistent with the predictions of gas-phase models of van Dishoeck and Black [5] and well correlated with those of CH (Fig. 7b).

First interstellar detection of $OH^+$ molecule in the ISM was possible while using radio observations [20]; the weak interstellar line at 3583.769 A was observed in several translucent sightlines by Krełowski et al. [21]. The radical was also detected in 2010 by the PRISMAS team using the Herschel space telescope in the far-IR range [22].

The NH hydrine has an electronic transition at 3358 Å which is accessible to the ground-based observations but very difficult to detect, as the features are very weak. It was detected in the ISM by Meyer and Roth [23]. Until this compilation the NH molecule was observed toward 10 diffuse and translucent sightlines by Weselak et al. [24] and references there in. Column densities of the NH molecule are related to CN and interstellar extinction. This result is consistent with the possible creation of NH by means of dust surface chemistry [25].

The first detection of interstellar CO molecule was possible using a rocket-borne spectrograph [26]; also other far-UV bands of the CO molecule were detected by Copernicus observers in many sightlines [27, 28]. Comparisons of $H_2$ and CO made by Liszt [29] and Sheffer et al. [30] showed a typical ratio $CO/H_2$ of about $10^{-6}$, the value which is consistent with the gas-phase models of van Dishoeck and Black [5]. Generally, the relation between column densities of CO and interstellar extinction is quadratic [31] and the CO molecule correlates more strongly with CN than with $H_2$ [32, 33, 34].

**Column densities**

To derive column density from an absorption line measurement one should obtain equivalent width of absorption line defined in terms of

$$W_\lambda = \int (F_c - F_\lambda)/F_c \, d\lambda \qquad (1)$$

where $F_\lambda$ is the measured flux level at wavelength $\lambda$ and where $F_c$ is the flux level in the continuum adjacent to the absorption line. The equivalent width is related to the optical depth $\tau(\lambda)$ according to the relation

$$W_\lambda = \int (1-\exp(-\tau(\lambda))) \, d\lambda \qquad (2)$$

which for low optical depths ($\tau(\lambda) \ll 1$) is expressed in the form

$$W_\lambda = \int \tau(\lambda) \, d\lambda = N \int \sigma(\lambda) \, d\lambda \qquad (3)$$

where N is the number of absorbers per $cm^2$, $\sigma(\lambda)$ is the absorption cross section.
In the case of low optical depths (it means that the spectral line formed in such conditions is not saturated) we can use the relation of Herbig [35] which gives proper column density when the absorption line is unsaturated,

$$N = 1.13 \, 10^{20} \, W_\lambda / (\lambda^2 f) \qquad (4)$$

where $W_\lambda$ and $\lambda$ are expressed in Angstroms and the column density in $cm^{-2}$.

To obtain column density we should also know oscillator strength f obtained in the aproximation of weak lines defined by Mulliken [36],

$$f = (4\varepsilon mc^2/e^2\lambda^2) \int \sigma(\lambda) \, d\lambda \tag{5}$$

where $\lambda$ is the central wavelength of the band, $\varepsilon$ is the vacuum permitivity, m and e are the mass and the charge of the electron and c is the speed of light.

Based on observations it is also possible to obtain correct oscillator strength values (f-values). The intensity ratio of two unsaturated spectral lines equals to:

$$\frac{W_{\lambda 1}}{W_{\lambda 2}} = \frac{f_1 \, (\lambda_1)^2}{f_2 \, (\lambda_2)^2} \tag{6}$$

Equation (6) describes the ratio of equivalent widths (as measured in spectrum) of two different lines of the same molecule in relation to their wavelengths and holds true if the lines arise in the same lower state and the lines are the same branch in both bands (for more information see Larsson 1983). For instance, based on equation (6) it was possible to obtain correct f-values of $CH^+$ (A-X) transitions as it was presented in the publication of Weselak et al. [37].

It is interesting to know that, in many cases, oscillator strengths obtained in the laboratory differ from those obtained on the basis of astronomical spectroscopy, ie. further experiments and/or observations seem necessary to determine this important property.
In Table 1 the oscillator strengths taken from the literature for known transitions of heteronuclear diatomics from near-UV to near-infrared wavelengths are presented.
Based on equation (4) it is possible to obtain precise column densities (besed on unsaturated lines) of heteronuclear diatomics which are presented in Table 2 toward selected diffuse and translucent targets taken from the literature.

**Column densities of heteronuclear diatomics and E(B–V)**

The color excess E(B–V) is the most popular measure of interstellar extinction (or reddening) and is expressed as

$$E(B-V) = (B-V) - (B-V)_0 = A_B - A_V \tag{7}$$

where $(B-V)_0$ is the intrinsic value of the colour index of the star, and $A_B$ and $A_V$ are total extinctions in the photometric B (4400 Å) and V (5500 A) bands, respectively.
Generally, extinction arises with distance (ie. with number of dust-type absorbers along the sight-line).
In Fig. 5 interstellar column densities of the CH, $CH^+$, CN and OH molecules are shown as being correlated with E(B-V). The CH is the most easily observable heteronuclear diatomic molecule which is closely related to $H_2$ [10, 31]; the interstellar $H_2$ is believed to be formed on dust grains.

**Relations between column densities of heteronuclear diatomics and molecular hydrogen**

Molecular hydrogen is the most abundant molecule in the ISM [38, 39].
Because it is symmetric and homonuclear it possesses no electric-dipole allowed vibrational or rotational transitions. As a result the only probes of $H_2$ molecule in the ISM are the far-UV electronic transitions below 1115 A [40].

The first spectrum of interstellar $H_2$ was acquired using rocket-based experiment toward ξ Per [41]. When, in 1972, the Copernicus orbital observatory was launched, it was possible to detect $H_2$ absorption toward more than 100 sightlines [42, 43].
The Far Ultraviolet Spectroscopic Explorer (FUSE) observatory launched in 1999 provides spectra toward additional sightlines [44]. Absorption lines of vibrationally excited molecular hydrogen lying in the mid-UV have been detected toward several objects using the spectrographs installed on the Hubble Space Telescope (HST). Federman et al. [45], detected $H_2$ absorption toward ζ Oph and Meyer [46] toward HD's: 37903, 37021, 37061, 147888.
In Fig. 6 it is presented a poor relation between column densities of $CH^+$ and $H_2$ molecules; also a weak relation is observed in case of $CH^+$ and OH.

The very good relations between column densities of CH and $H_2$ molecules and also between column densities of OH and CH are presented in Fig. 7. According to the very good relation between column densities of CH, $H_2$ and OH diatomics, the easily observed CH molecule may be used as a good tracer of interstellar $H_2$ and OH which are difficult to observe.

**Heteronuclear diatomics and DIBs**

The first two diffuse interstellar bands (DIBs) were discovered by Heger [47]. The application of solid state detectors to DIB observations led to discoveries of new features. Currently, the list of known DIBs contains 414 entries ([48]) a majority of them is very shallow. Even more importantly, the fine structure (reminiscent of the rotational contours of bands of polyatomic molecules) has been detected in some DIBs [49, 50]. Nearly all conceivable forms of matter - from hydrogen anion to dust grains - have already been proposed as DIB carriers, so far with no generally accepted success.
The spectral region, where DIBs occur (from visible to near infrared), is characteristic for electronic transitions of complex chemical species. The nearly constant positions and nearly unchanged profiles of DIBs, irrespective of the line of sight, point to free gas-phase molecules as the carriers of these bands. Among the proposed carriers the most likely ones are: (hydro)carbon chains, Polycyclic Aromatic Hydrocarbons (PAHs) and fullerenes. High resolution, high S/N profiles of DIBs may play a decisive role in identification of their carriers by means of comparison with laboratory, gas-phase spectra. The list of strong DIBs, detected in the spectrum of HD 183143, with the values of equivalent widths higher than 100 mÅ is presented in Table 3 of this chapter.
In Fig. 8 it is presented region of two strong 5780 and 5797 DIBs in spectrum of of highly reddened star HD 168607.

The abundance of elements in the Universe constrains the chemical composition of the carriers of DIBs. They ought to be built out of the most abundant elements: H, O, C, N, a small contribution of other elements also cannot be excluded. Among this group of elements carbon is in an exceptional position, since it can form a great number of stable compounds with a linear, planar and spherical structure. Bare carbon chains of 5 to 15 C atoms have been proposed as carriers of diffuse interstellar bands by Douglas [51], Fulara et al. [52] and Freivogel et al. [53] later extended this hypothesis to the whole class of linear, unsaturated hydrocarbons. The species trapped in neon matrices, according to the mass selection of $C_nH_m$- showed strong absorption bands in the red and in the near infrared regions of the spectrum. The above suggestion seems thus attractive. The interstellar medium contains a lot of molecules based on carbon chain skeletons. Their vibrational and electronic transitions cause features in the spectral range full of diffuse bands. Apart from (hydro)carbon chains and PAHs, also fullerenes (recently identified in circumstellar shells by Cami et al. [9]) fall in this category having features in visible spectral range.
The origin of DIBs is still not precisely known; possibly spectra of many carriers are covering the whole spectral region in which they exist. Discussion on the possible carriers can be found in publications of Herbig [54], Sarre [55], Oka and McCall [56].

Correlation analysis between column densities of simple diatomics and observed DIB intensities may shed some light on the origin of the latter. Since the publication of Weselak et al. [57] it is well seen that carriers of the 5780 and 5797 DIBs are better correlated with CH than with CN molecule. Actually it is known that strong DIBs correlate well with each other [58] and intensity of the strong 5780 DIB is better correlated with atomic than molecular hydrogen. The best correlation is known in the case of 6196 and 6614 DIBs ([59], [60]).

The recent suggestion that profile widths of (some?) DIBs are related to rotational temperatures of simple carbon species [61] needs a special attention. If DIB carriers are centrosymmetric, ubiquitous molecules, they should be present in the proposed chains of chemical reactions.

**Conclusions**

The above considerations allow us to infer the following conclusions:

1. Heteronuclear diatomics play an important role in understanding physical conditions and chemistry of diffuse and translucent clouds they do populate. They form electronic spectra which cover the spectral region from near-UV to infrared.
2. Based on observed spectra it is possible to obtain precise values of equivalent widths; knowing precise oscillator strengths it is possible to obtain molecular column densities toward observed targets. These column densities may be compared with those modelled (using e.g. the PDR Meudon code).
3. Interstellar molecules such as $H_2$, CH and OH seem to be closely connected in translucent clouds. Other molecular species, such as the CN or the CH cation seem to be formed in other reaction pathways since their column densities do not correlate with those of $H_2$, CH and OH.
4. An analysis of correlations between column densities of diatomic molecules and DIB intensities may shed some light on nature of the carriers of the latter as well as on physics and chemistry of the environments in which they are formed.

It is emphasized, that the future observational work should be focused on spectra of high quality of objects situated behind diffuse and translucent clouds to allow additional analysis of the observed transitions of heteronuclear diatomics.

**Acknowledgements**

This paper was supported by the Polish State Committee for Scientific Research under grant 5820/B/H03/2011/40.

**Table 1.** List of heteronuclear diatomics observed in the near-infrared, visible and near-UV spectral region toward translucent sightlines.
References are as follows: (1) – Roth & Meyer [62], (2) – van Dishoeck & Black [63], (3) – Lien [66], (4) – Weselak et al. [65], (5) – Lien [66], (6) – Weselak et al. [37], (7) – Weselak et al. [68], (8) – Krełowski et al. [21], (9) – Weselak et al. [24].

| Molecule | Band | Line | $\lambda$ | f-value | References |
|---|---|---|---|---|---|
| CN | B–X (0–0) | R(0) | 3874.608 | 0.0342 | (1) |
| | | R(1) | 3873.998 | 0.0228 | |
| | | P(1) | 3875.763 | 0.0114 | |
| | | R(2) | 3873.369 | 0.0205 | |
| | | P(2) | 3876.310 | 0.0137 | |
| | B–X (1–0) | R(0) | 3579.963 | 0.0030 | (1) |
| | | R(1) | 3579.453 | 0.0020 | |
| | | P(1) | 3580.937 | 0.0010 | |
| | A–X (1–0) | $R_1(0)$ | 9186.935 | 0.000792 | (2) |
| | | $^RQ_{21}(0)$ | 9144.043 | 0.000501 | |
| | | $^SR_{21}(0)$ | 9139.677 | 0.000207 | |
| | | $^QR_{12}(1)$ | 9190.110 | 0.000702 | |
| | | $Q_2(1)$ | 9147.201 | 0.000501 | |
| | | $Q_1(1)$ | 9190.120 | 0.0003345 | |
| | | $R_2(1)$ | 9142.833 | 0.000297 | |
| | | $R_1(1)$ | 9183.216 | 0.000501 | |
| | A–X (2–0) | $R_1(0)$ | 7906.601 | 0.00040128 | (2) |
| | | $^RQ_{21}(0)$ | 7874.847 | 0.00025384 | |
| | | $^SR_{21}(0)$ | 7871.643 | 0.00010488 | |
| | | $^QR_{12}(1)$ | 7908.952 | 0.00035644 | |
| | | $Q_2(1)$ | 7877.189 | 0.00025384 | |
| | | $Q_1(1)$ | 7908.959 | 0.00017024 | |
| | | $R_1(1)$ | 7903.896 | 0.00025384 | |
| CH | A–X (0–0) | $R_{2e}(1) + R_{2f}(1)$ | 4300.3132 | 0.00506 | (2) |
| | B–X (0–0) | $R_2(1)$ | 3878.774 | 0.00110 | (3) |
| | | $Q_2(1) + {}^QR_{12}(1)$ | 3886.409 | 0.00320 | |
| | | $^PQ_{12}(1)$ | 3890.217 | 0.00210 | |
| | B–X (1–0) | $R_2(1)$ | 3627.403 | 0.00035 | (4) |
| | | $Q_2(1) + {}^QR_{12}(1)$ | 3633.289 | 0.00104 | |
| | | $^PQ_{12}(1)$ | 3636.222 | 0.00069 | |
| | C–X (0–0) | $R_2(1)$ | 3137.576 | 0.00021 | (5) |
| | | $Q_2(1) + {}^QR_{12}(1)$ | 3143.183 | 0.00054 | |
| | | $^PQ_{12}(1)$ | 3145.996 | 0.00043 | |
| CH$^+$ | A–X (0–0) | R(0) | 4232.548 | 0.00545 | (6) |
| | A–X (1–0) | R(0) | 3957.689 | 0.00342 | |
| | A–X (2–0) | R(0) | 3745.308 | 0.00172 | |
| | A–X (3–0) | R(0) | 3579.024 | 0.00075 | |
| | A–X (4–0) | R(0) | 3447.077 | 0.00040 | |
| OH | A–X (0–0) | $Q_1(3/2)+{}^QP_{21}(3/2)$ | 3078.443 | 0.00105 | (7) |
| | | $P_1(3/2)$ | 3081.6645 | 0.000648 | |
| OH$^+$ | A–X (0–0) | $R_{11}(0)$ | 3583.769 | 0.0030764 | (8) |
| NH | A–X (0–0) | $^RQ_{21}(0)$ | 3353.924 | 0.00240 | (9) |
| | | $R_1(0)$ | 3358.053 | 0.00410 | |

**Table 2.** Observed molecular column densities (in $10^{12}$ cm$^{-2}$) of heterodiatomics toward selected targets (most of data taken from Weselak et al., [24]) based on unsaturated lines. Data on CO molecule were taken from the publication of Federman et al. [67] and OH$^+$ from Krełowski et al. [21].

| HD number | Sp/L | E(B-V) | N(CN) | N(CH) | N(CH$^+$) | N(OH) | N(OH$^+$) | N(NH) | N(CO) |
|---|---|---|---|---|---|---|---|---|---|
| 23180 | B1III | 0.27 | 1.82 | 22.24 | 7.89 | 78.00 | | | 1096 |
| 24398 | B1Iab | 0.29 | 3.12 | 22.30 | 4.28 | 40.50 | | 0.90 | 1202 |
| 27778 | B3V | 0.37 | 26.76 | 39.93 | 10.55 | 102.00 | | 2.70 | 25118 |
| 147889 | B2III/IV | 1.00 | 25.06 | 100.33 | 33.90 | 252.36 | | 5.85 | |
| 149757 | O9.5V | 0.28 | 2.68 | 25.44 | 29.32 | 40.98 | | 0.76 | 2290 |
| 151932 | WN7 | 0.50 | 2.43 | 26.76 | 15.13 | 79.15 | 3.2 | | |
| 152270 | WC7 | 0.50 | 1.59 | 17.79 | 25.10 | 56.54 | | | |
| 154368 | O9Ia | 0.80 | 26.44 | 61.75 | 21.46 | 169.84 | 3.3 | 6.92 | 12302 |
| 154811 | O9.5Ib | 0.66 | 0.17 | 22.85 | 39.43 | 51.90 | | | |
| 163800 | O7 | 0.56 | 2.96 | 34.36 | 16.02 | 78.43 | 3.6 | 2.18 | |
| 164794 | O4V | 0.36 | 0.27 | 11.86 | 10.32 | 32.26 | | | |
| 169454 | B1Ia | 1.10 | 40.22 | 42.90 | 19.50 | 98.76 | | 6.86 | |

**Table 3.** List of selected strong Diffuse Interstellar Bands with equivalent width $W_\lambda > 100$ mA observed toward star HD 183143.
References are as follows: (1) – Herbig [54], (2) – Hobbs et al. [48]

| Wavelength [A] | $W_\lambda$ [mA] | FWHM [A] | Reference - |
|---|---|---|---|
| 4066 | 350 | 15 | (1) |
| 4180 | 700 | 25 | (1) |
| 4428.83 | 5700 | 22.56 | (2) |
| 4501.66 | 211.2 | 3.01 | (2) |
| 4727.16 | 156.2 | 3.11 | (2) |
| 4762.62 | 126.5 | 2.5 | (2) |
| 4881.06 | 343.7 | 11.31 | (2) |
| 5361.13 | 118.4 | 7.15 | (2) |
| 5420.01 | 166.4 | 12.36 | (2) |
| 5450.52 | 359.5 | 12.30 | (2) |
| 5487.7 | 235.8 | 3.38 | (2) |
| 5508.42 | 158.8 | 2.66 | (2) |
| 5535.82 | 139.1 | 3.91 | (2) |
| 5705.31 | 172.5 | 2.68 | (2) |
| 5705.92 | 279.8 | 14.22 | (2) |
| 5780.61 | 779.3 | 2.14 | (2) |
| 5797.20 | 186.4 | 0.91 | (2) |
| 5843.92 | 100.7 | 4.38 | (2) |
| 6010.65 | 202.8 | 4.01 | (2) |
| 6177 | 2390 | 29 | (1) |
| 6203.14 | 206.2 | 1.42 | (2) |
| 6205.20 | 151.3 | 2.47 | (2) |
| 6269.93 | 256.4 | 1.3 | (2) |
| 6284.28 | 1884.2 | 4.02 | (2) |
| 6307 | 220 | 9 | (1) |
| 6318 | 210 | 8 | (1) |
| 6379.32 | 105.4 | 0.68 | (2) |
| 6413.17 | 118.2 | 8.28 | (2) |
| 6494.14 | 454.8 | 9.20 | (2) |
| 6534.54 | 329.2 | 12.69 | (2) |
| 6613.70 | 341.6 | 1.08 | (2) |
| 6940 | 410 | 18 | (1) |
| 6993.24 | 173.1 | 0.94 | (2) |
| 7224.18 | 358.8 | 1.24 | (2) |
| 7405.5 | 117 | 15 | (1) |
| 7429 | 560 | 21 | (1) |
| 7562.29 | 100.5 | 1.72 | (2) |
| 7927 | 360 | 14 | (1) |
| 8038.58 | 204.7 | 5.37 | (2) |
| 8620.99 | 355.8 | 3.57 | (2) |
| 8649 | 370 | 14 | (1) |
| 9577 | 470 | 4 | (1) |
| 9632 | 780 | 1 | (1) |
| 11797.5 | 170 | 2.7 | (1) |
| 13175 | 420 | 4 | (1) |

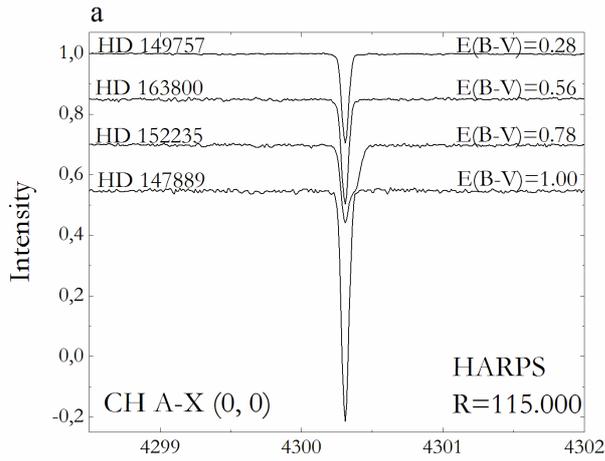

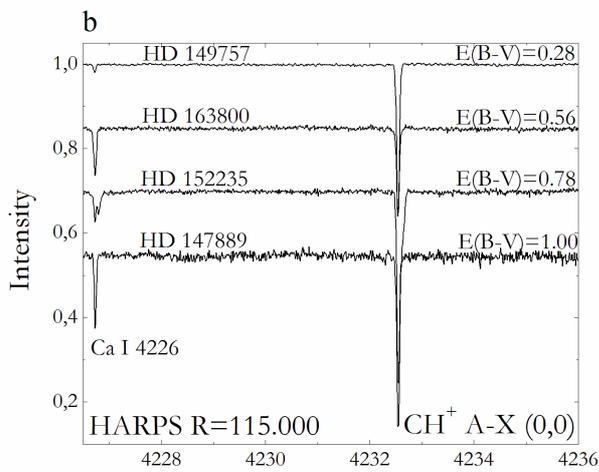

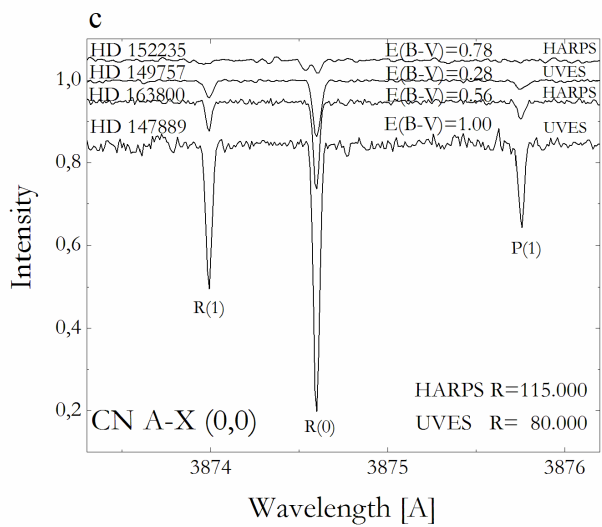

Fig. 1. The CH A-X system at 4300 A in spectra of four targets with the increasing reddening (Panel a - from top to bottom). In Panel b we present $CH^+$ A-X (0, 0) band at 4232 A seen in the spectra of the same stars. In the case of HD 152235 additional Doppler component is seen in $CH^+$ and CaI profiles. Panel c presents the CN B-X (0, 0) band centered near 3875 A. The Doppler splitting in the spectrum of HD 152235 is evident in the CN B-X system (spectrum at the top of this figure). Note also low intensity of CN features as compared to those of CH or $CH^+$ in the spectrum of this star.

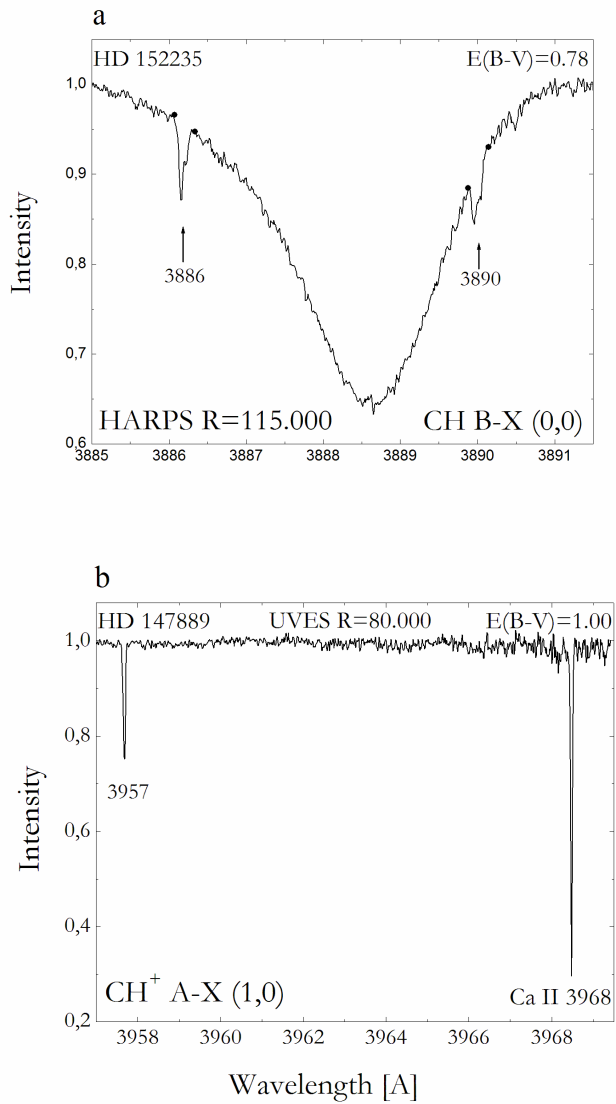

Fig. 2. The CH B-X features centered near 3886 and 3890 A seen in the spectrum of HD 152235. Position of the continuum is marked with dots (Panel a). It is also presented $CH^+$ A-X (1, 0) band near 3957 A in the spectrum of highly reddened HD 147889 - Panel b. These features are used in many cases to determine column densities when the bands of CH or $CH^+$ seen in Fig. 1, are saturated.

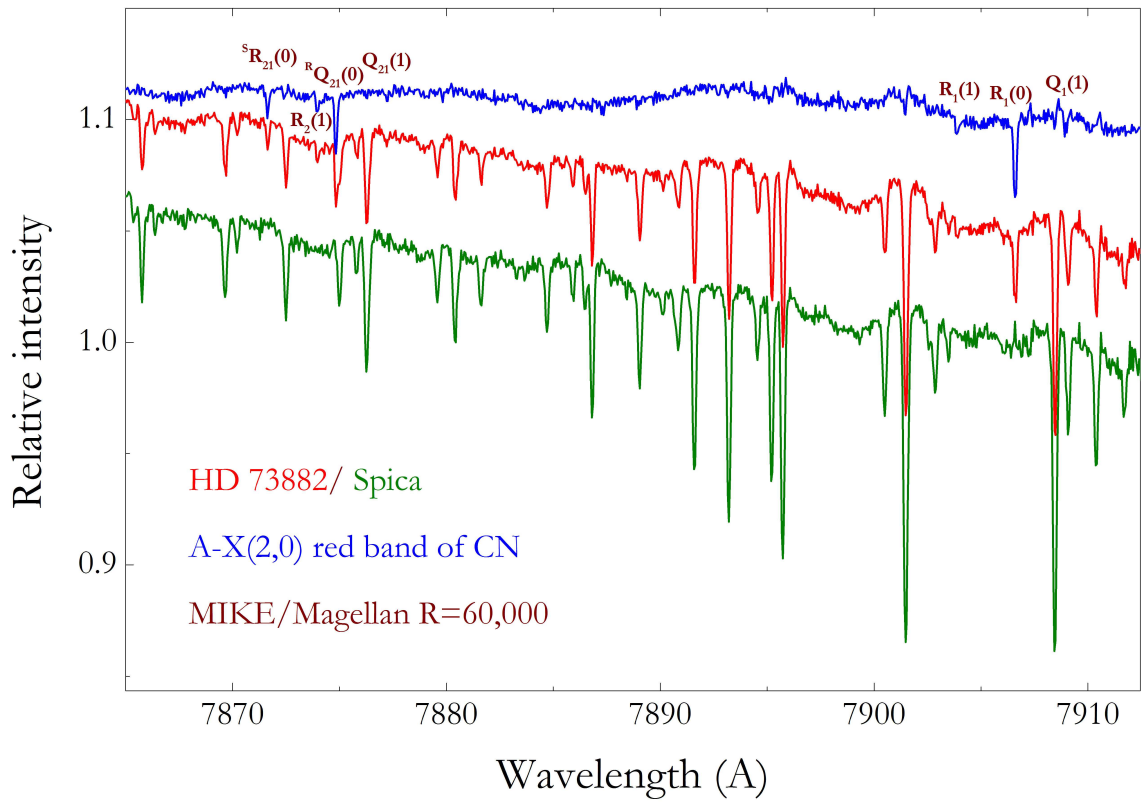

Fig. 3. Spectrum of HD 73882 after removal of telluric lines (at the top) in the region of CN A-X (2, 0) transition. Spectrum of divisor (Spica) is presented with red line in the centre. This figure is shown by courtesy of G.A. Galazutdinov and J. Krełowski.

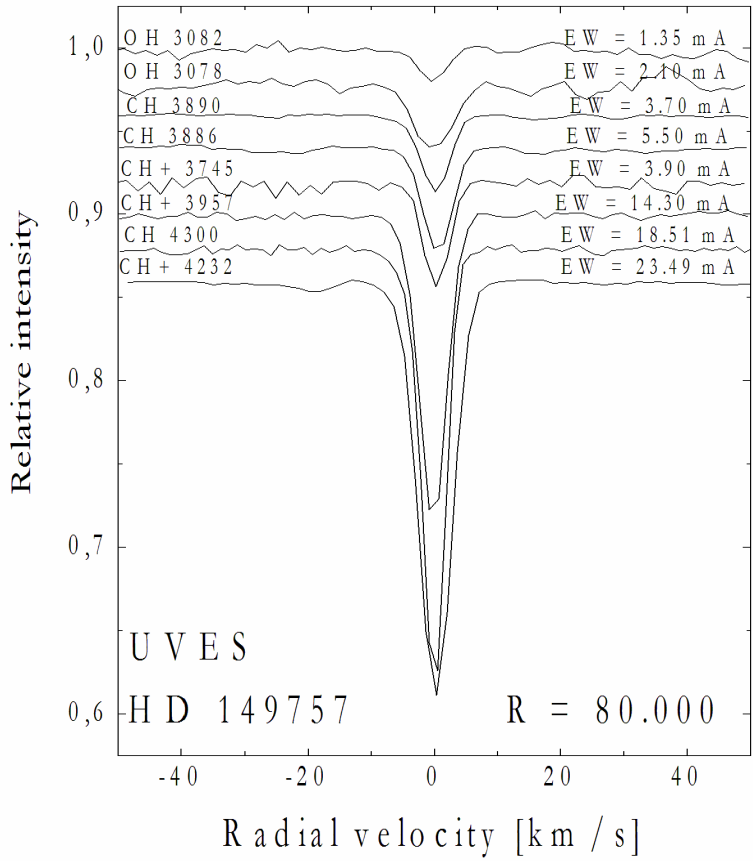

Fig. 4. Interstellar features of the OH, CH and CH$^+$ molecules (presented in Table 1) in the spectrum of HD 149757. The CH$^+$ line at 4232 A is saturated (saturation effects are well seen when EW > 20mÅ – see Weselak et al. [57]). In this case to obtain the column density one should use the unsaturated CH$^+$ line at 3957 A.

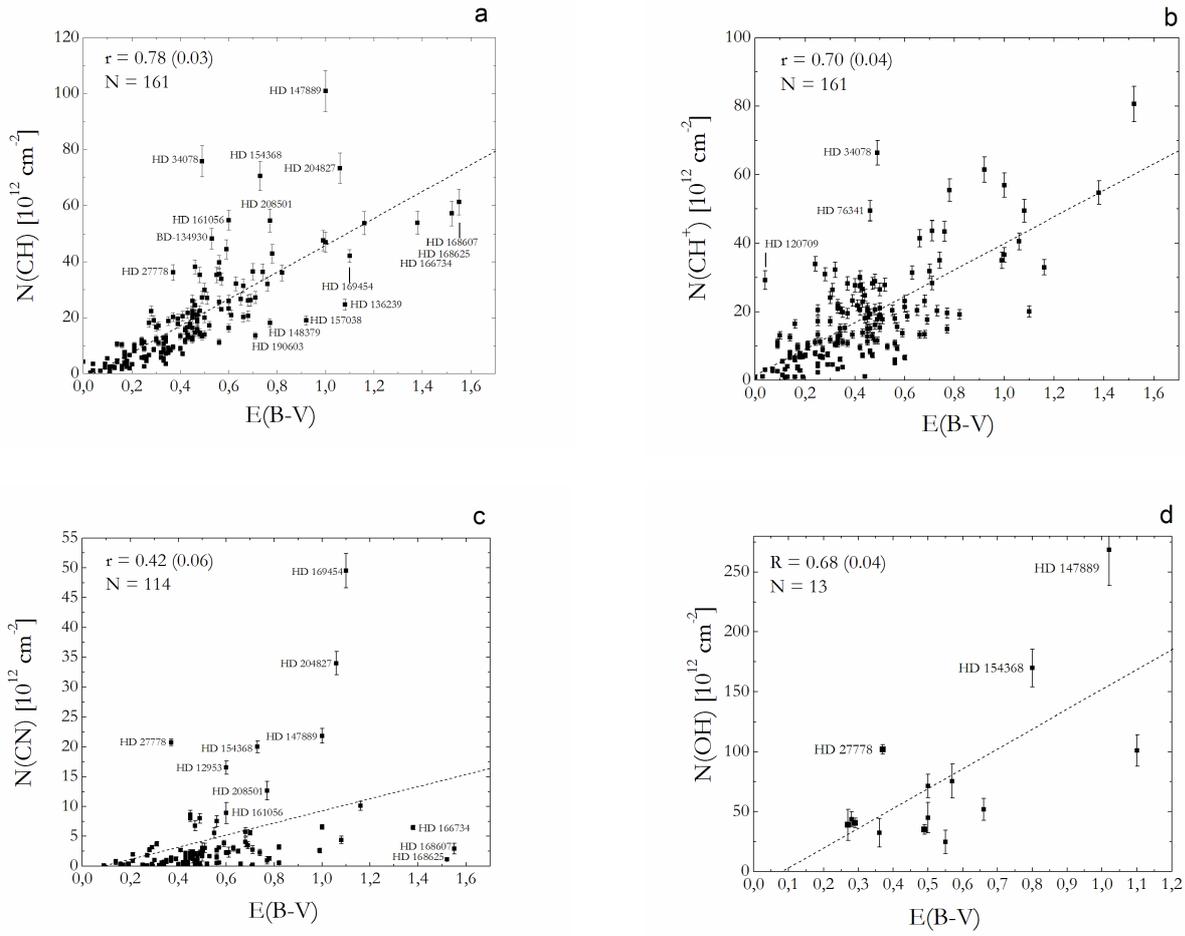

Fig. 5. Column densities of CH, CH$^+$, CN and OH correlated with E(B-V) . In each case the best fit is plotted with dash line.

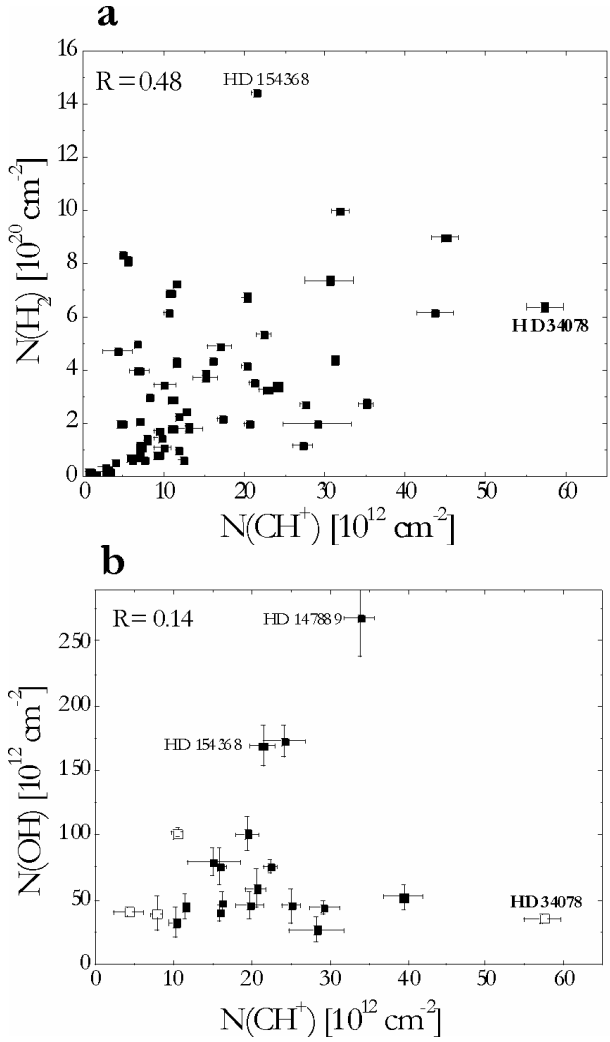 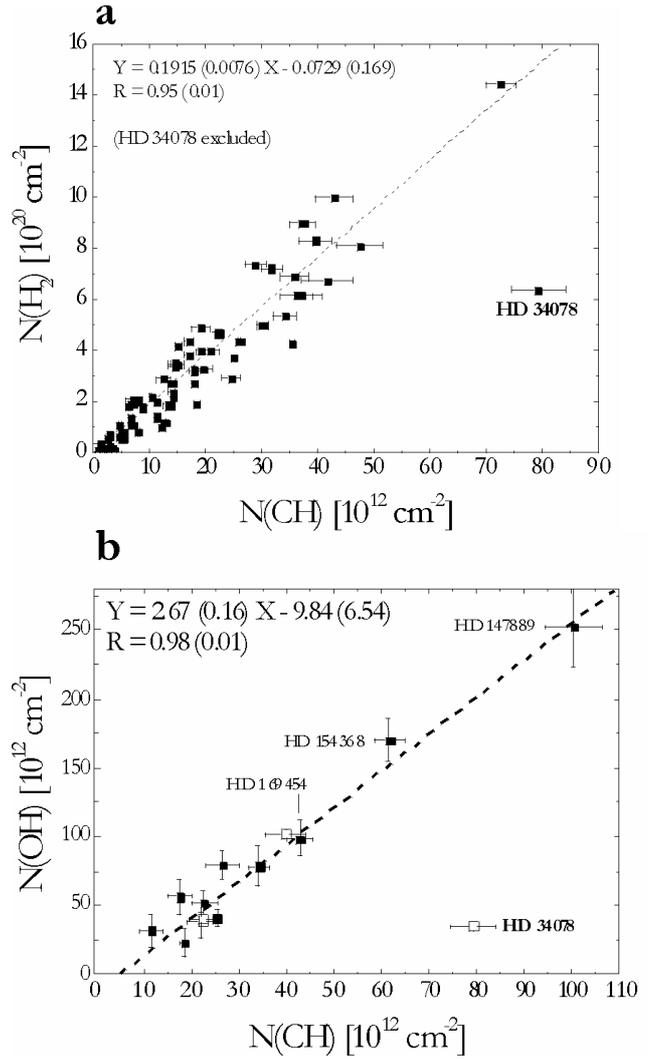

Fig. 6. The poor relations between column densities of $CH^+$ and $H_2$ molecules (a) and also between the column densities of OH and $CH^+$ (b) data on OH taken from Weselak et al. [17]. The open squares mark objects where column densities of the OH molecule were taken from the literature. Correlation coefficients are presented at the top-left in each case.

Fig. 7. The very good relations between column densities of $H_2$ and CH molecules (a) and also between the column densities of OH and CH (b) – values based on the publication of Weselak et al. [17]. With the dashed line we present the linear relation with one data-point excluded (HD 34078). With open squares we mark objects where column densities of the OH molecule were taken from the literature.

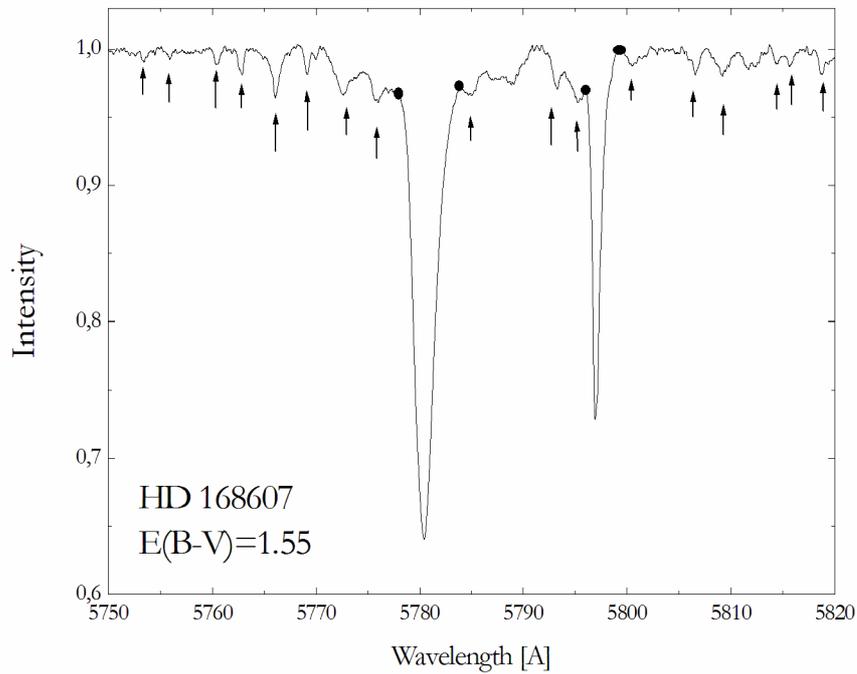

Fig. 8. Strong 5780 and 5797 diffuse bands in the spectrum of highly reddened star HD 168607. With dots we mark the continuum level used to obtain the equivalent widths of 5780 and 5797 DIBs blended with broad and weak DIBs at 5778 and 5795 A. Many weak DIBs are also depicted with arrows. Spectrum from the UVES instrument acquired with resolution R=80.000.